\documentclass[cits]{PoS}
\usepackage[square,comma,numbers]{natbib}
\usepackage{graphicx}
\usepackage{sidecap}

\newcommand{\uv}{\mbox{$u$-$v$}}

\newcommand{\Jb}{\mbox{Jy bm$^{-1}$}}
\newcommand{\thout}{\mbox{$\theta_{\rm out}$}}
\newcommand{\thin}{\mbox{$\theta_{\rm in}$}}
\newcommand{\arcmin}{$^{\prime}$}

\title{Radio Imaging of SN 1993J: The Story Continues}

\ShortTitle{Radio Imaging of SN 1993J}

\author{\speaker{Michael Bietenholz}\thanks{Also at York University, Toronto, Canada.}\\
        Hartebeesthoek Radio Observatory, Krugersdorp, South Africa\\
        E-mail: \email{michael@hartrao.ac.za}}

\author{N. Bartel, {\small\it York University, Toronto, Canada}; 
M. P.  Rupen, {\small\it NRAO, Socorro, NM, USA};
V. V. Dwarkadas {\small\it U. Chicago, USA};
A. J. Beasley, {\small\it NEON, Boulder, Colorado, USA};
D. A. Graham, {\small\it MPIfR, Bonn, Germany};
T. Venturi, {\small\it IRA, INAF, Italy};
G. Umana, {\small\it INAF, Italy};
W. Cannon, {\small\it York University, Toronto, Canada}; 
J. Conway {\small\it Onsala Space Observatory, Onsala, Sweden}}

\abstract{We present the most recent VLBI images of SN 1993J, taken at
1.7 GHz on 2010 March 5-6, along with a discussion of its evolution
with time.  The new image is the latest in a sequence
covering almost the entire lifetime of the supernova.  For these
latest observations we used an "in beam calibrator" technique, and
obtained a background rms brightness of 3.7 $\mu$\Jb.  The supernova
shell remains quite circular in outline.  Modulations in brightness
are seen around the rim which evolve relatively slowly, having
remained generally similar over the last several years of observation.
We determine the outer radius of the supernova using visibility-plane
model-fitting.  The supernova has slowed down to around 30\% of its
original expansion velocity, and continues to expand with radius
approximately $\propto t^{0.8}$, however, deviations from a strict
power-law evolution are seen.  We do not find any clear-cut evidence
for systematically frequency-dependent evolution, suggesting that the
radii as determined from visibility-plane model-fitting continue to
provide reasonable estimates of the physical outer shock-front radius.

}

\FullConference{10th European VLBI Network Symposium and EVN Users Meeting: VLBI and the new generation of radio arrays\\
		September 20-24, 2010\\
		Manchester Uk}

\begin{document}

\section{Introduction}

SN~1993J, in the nearby galaxy M81, was one of the nearest as well as
one of the radio-brightest core-collapse supernovae observed in recent
history \citep[e.g.,][]{SN93J-Sci, SN93J-1, SNVLBI-Bologna}. We
undertook an extensive campaign of VLBI imaging (see \citep{SN93J-3}
and references therein; note that a second, parallel campaign of VLBI
observations was also undertaken, see \citep{Marti-Vidal+2011a} for
the most recent results from this second campaign).
One of the results of our VLBI campaign was in fact a direct,
geometrical estimation of the distance to SN~1993J and M81 using the
expanding shock front method, which resulted in a value of $3.96 \pm
0.29$~Mpc.  This value is consistent with that obtained from Cepheid
measurements by Huterer et al.\ \citep{HutererSS1995}, but
$9\pm13$\% larger than the more commonly cited value of Freedman et
al.\ \citep{Freedman+2001}, (see \citep{SN93J-4} for a fuller
discussion).  In this paper, we briefly present some of the most
recent developments in our continuing VLBI campaign.

\begin{SCfigure}
\includegraphics[width=0.7\textwidth]{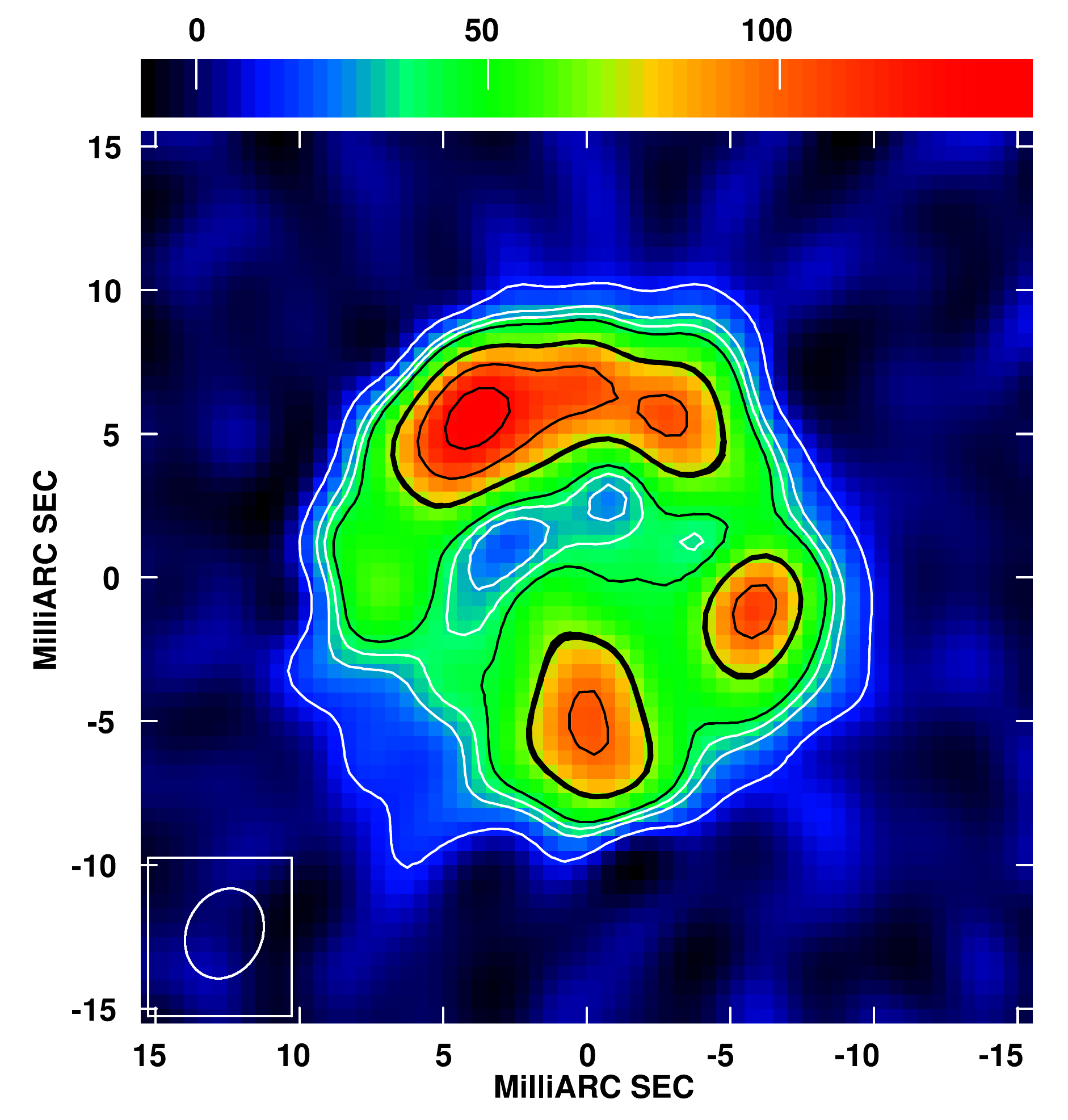} 
\caption{VLBI image of SN~1993J at 1.7 GHz, taken on 2010 March 5-6,
or at $t = 16.9$~yr after the explosion. Contours are at 10, 20, 30,
40, 50, 70 and 90\% of the peak brightness of 117~$\mu$\Jb, and the
rms background brightness was 3.7~$\mu$\Jb.  The FWHM of the
convolving beam is indicated at lower left.  North is up and east is
to the left.}
\label{fmr10image}
\end{SCfigure}

Our latest VLBI image was obtained on 2010 March 5 - 6, using a global
array of 18 telescopes from the European VLBI Network and the
National Radio Astronomy Observatory.  We observed at a frequency of
1.7~GHz, with a total time of 24 hours.  As throughout our campaign,
we used M81*, the active core of M81, as a phase reference source
\citep{M81-2004}.  Since SN~1993J and M81\* are only 3\arcmin\ apart
and are within the same primary beam of all the
telescopes\footnote{We did not use the data from the phased Westerbork
array in these preliminary results, as the two sources are not within
the synthesized beam of the phased array. We defer a fuller discussion
of results including the Westerbork data to a future publication.}
at 1.7~GHz, we performed the phase-referencing using an in-beam
calibrator technique where the telescopes were pointed at a location
between the two sources.  The total CLEAN flux density recovered for
SN~1993J was 1.7~mJy, while the rms background brightness was
3.7~$\mu$\Jb.
We show the image of SN~1993J in Figure~\ref{fmr10image}.  Although
the dynamic range in the image was modest (peak/rms = 32),
the main features of the image are quite consistent with those seen
earlier and at other wavelengths.  The outline of the emission region
remains highly circular.  Along the ridge line, we see a broad
enhanced brightness just east of north, and a smaller one just south
of west, characteristics which are already apparent in the images from
2001 and 2002 \citep{SN93J-3}.  We defer a fuller discussion of the
deviations from uniform emissivity around the ridge-line to a future
paper.  There is no central brightening, in contrast to what is seen
in SN~1986J, where a prominent central brightening may mark the
appearance of a young pulsar-wind nebula \citep{SN86J-2}.

In addition to producing an image, we also determined precise values
for the supernova's inner and outer angular radii, \thin\ and \thout\
respectively, by fitting an optically-thin spherical shell
model\footnote{The other free parameters in our model are the center
position and the total flux density of the supernova}
directly to the visibility measurements \citep[see][]{SN93J-2}.
The Chevalier mini-shell model \citep{Chevalier1982b} for a supernova
predicts that its radius should evolve with time in a power-law
fashion, provided the densities in both the circumstellar medium (CSM)
and the ejecta are also power-law functions of radius.  In this
model, the radius of the supernova's shock-front is $\propto
t^m$, where $t$ is time, and $m$ is the power-law index.  SN~1993J has
indeed, over its history, shown an evolution quite close to such a
canonical power-law (see, e.g., \citep{SN93J-2, Marti-Vidal+2011a}),
with an average value of $m \sim 0.8$.  However, the measurements of
SN~1993J are of sufficient quality that relatively small deviations
from a power-law, or changes in time of the value of $m$, can be
determined.  In particular, both changes in the lightcurve decay and
in the expansion index show that the evolution of SN~1993J is not
self-similar. 

\begin{SCfigure}
\includegraphics[width=0.64\textwidth]{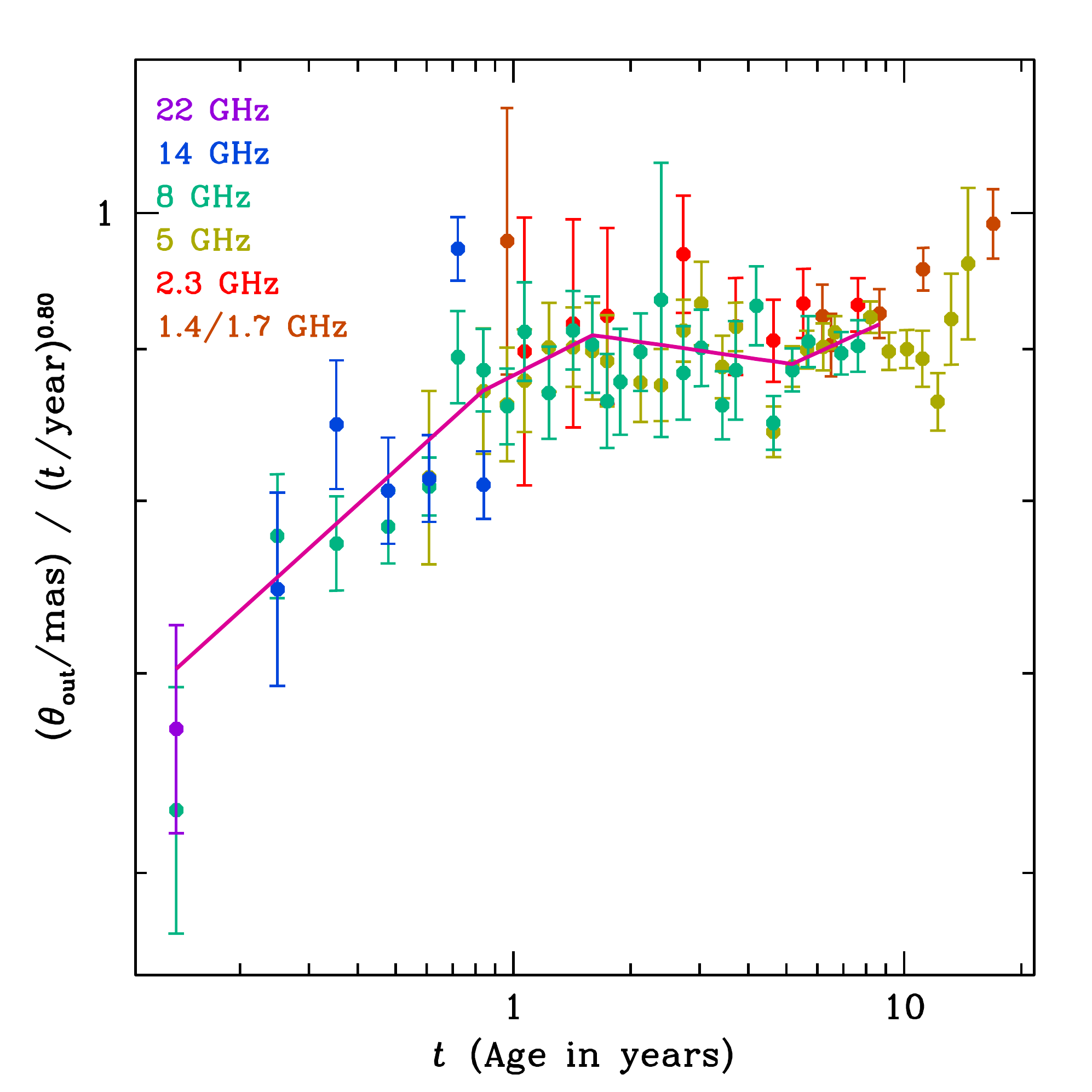}
\caption{The scaled outer angular radius of SN~1993J as a function of
time.  The scaled angular radii are ${\thout / {(\rm mas)}} \over
{(t/{\rm yr})^{0.8}}$. In other words we plot the deviation of the
measured angular radii from a simple power-law expansion with $\theta
\propto t^{0.8}$.  This scaling of the measured radii highlights the
observed deviations from a simple power-law expansion.  The points
show the measured outer radii along with their uncertainties, with
different colours representing measurements at different frequencies.
The purple broken line indicates the fit of Bartel et al.,
\citep{SN93J-2} to the measurements up to $t = 8$~yr.  For $t <
2.7$~yr, a ratio of $\thout/\thin = 1.25$ was assumed, while for later
times \thin\ was fit simultaneously with \thout\
\citep[see][]{SN93J-2, SN93J-4}.}
\label{frovstplot}
\end{SCfigure}

In Fig.~\ref{frovstplot} we compare the evolution of the outer angular
radii of SN~1993J to a canonical power-law with $\theta \propto
t^{0.8}$.  The most clear break from a single power-law expansion
curve is seen in at $t \simeq 1$~yr, after which a distinct increase
in the deceleration occurs.  Smaller changes in the slope of the
power-law expansion occur at later times.

\begin{SCfigure}[][t]
\includegraphics[width=0.58\textwidth]{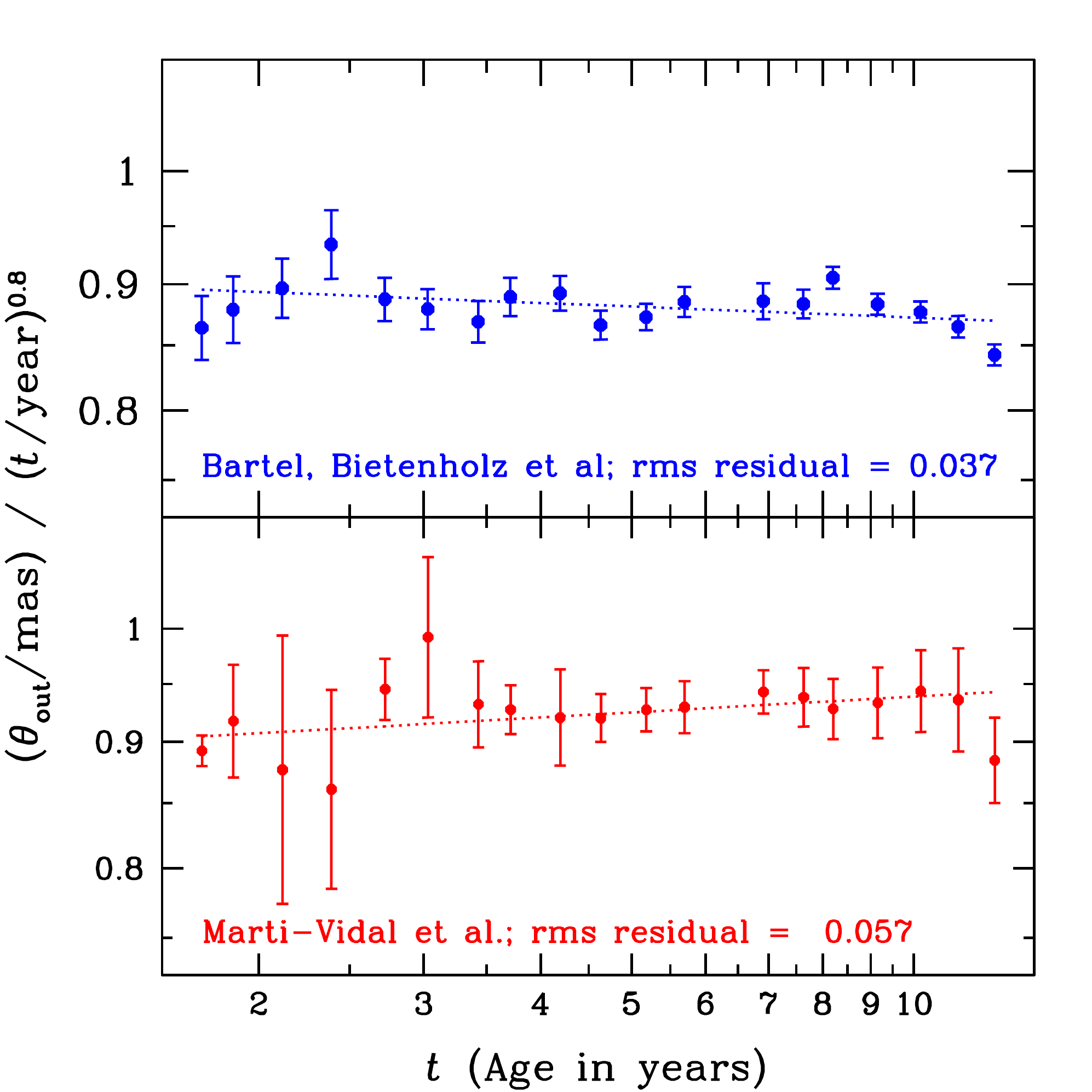}
\caption{Comparison of the outer angular radii determined using
\uv~plane model-fitting with those obtained using the image-plane CPM
(``common-point'' method) by \citep{Marti-Vidal+2011a}.  We plot the
published values with uncertainties for only those epochs of our
observations for which radii have been determined by both methods and
for which the supernova is large enough to apply the CPM (i.e., $t >
1.6$~yr). As in Fig.~\protect\ref{frovstplot} above, we plot the ratio
${\thout / {(\rm mas)}} \over {(t/{\rm yr})^{0.8}}$, or the radii as
scaled by nominal power-law expansion with $\theta \propto t^{0.8}$.
In each panel, the dotted lines indicate a weighted least-squares fit
of a simple power-law to the plotted data, and the rms residual to
this fit (in unitless scaled radii) is indicated.  {\bf Top:} our
values \citep{SN93J-2}, obtained using \uv~plane model fitting.  {\bf
Bottom:} the CPM values of \citep{Marti-Vidal+2011a}. }
\label{fusvsthem}
\end{SCfigure}

We note here that Marti-Vidal et al.\ \citep{Marti-Vidal+2011a} prefer
determining the supernova angular radius in the image plane using a
method they call the ``Common Point Method'' or CPM
\citep{Marcaide+2009b}.  They re-reduced our earlier observations and
determined radii at these epochs using CPM, which they claim produces
results superior to those obtained from \uv~plane model-fitting.  We
find however, that a comparison of CPM values determined by
Marti-Vidal et al.\ for our observations with our originally published
radii, determined by \uv~plane fitting, with does not support this
claim.  Although the agreement between the CPM and our \uv~plane
radius values is generally good, our original values show
significantly less scatter than those determined using CPM\@. In
Fig.~\ref{fusvsthem}, we compare Marti-Vidal et al.'s determinations
of the radius with our own for those epochs for which both
determinations are available.  Our original values show significantly
lower scatter with respect to a smooth power-law evolution as well as
having smaller uncertainties.  We therefore continue to use \uv~plane
model-fitting as the method of choice for determining accurate values
for the angular radius of SN~1993J as a function of time.  We note
also both \uv~plane model-fitting and the CPM produce values of the
radius that are model-dependent in the sense that they can only be
precisely related to a physical radius by assuming some particular
form for the brightness profile of the supernova.

The measured angular radii at different frequencies do not always
agree within the experimental uncertainties, and we note a tendency
for the radii determined from data at lower frequencies to be slightly
larger than those from higher frequencies.  Over the period between $t
= 8$~yr and the last observations, the scaled radius at 1.7~GHz was,
on average, 5.7\% $\pm$ 5.2\% larger than that at frequencies $>
4$~GHz.  Our values, however, do not suggest any clearly systematic
frequency-dependence in the evolution of the radius as claimed by
other authors \citep{Marti-Vidal+2011a, Marcaide+2009b}.  We rather
think that small time- and frequency-dependent deviations of the true
brightness distribution from our geometrical model are likely
responsible for the small variations of the fitted radii with
frequency.  Since our determinations show no clear frequency
dependence, we think that the radii as determined by \uv~plane
model-fitting continue to provide a reasonable estimate of the outer
shock front radius.

\begin{SCfigure}[][!h]
\includegraphics[width=0.58\textwidth]{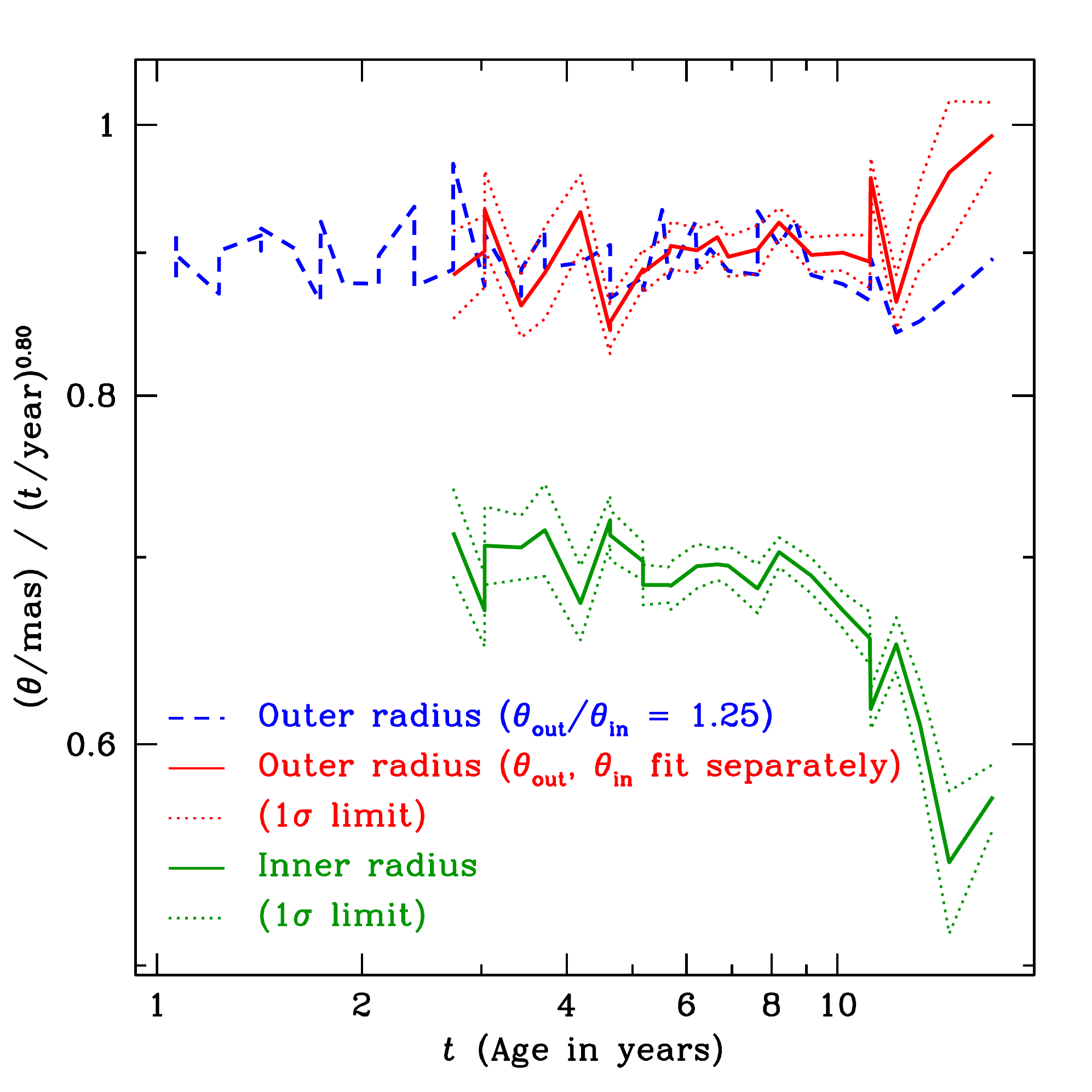}
\caption{A plot showing the evolution of the outer and inner angular
radii of SN~1993J as a function of time.  Again, as in
Figs.~\protect\ref{frovstplot} and \protect\ref{fusvsthem} above, we
plot the radii as scaled to a nominal power-law expansion with $\theta
\propto t^{0.8}$, in order to highlight any deviations from a simple
power-law evolution.  The red and green solid lines show the outer and
inner radii, respectively, when the two are fit separately.  For
comparison, we also show in blue the outer radius obtained when using
a fixed assumed ratio of $\thout/\thin = 1.25$.  Note that for $t <
2.7$~yr, the resolution was insufficient to reliably determine \thin\
separately, and we determined \thout\ by assuming $\thout/\thin =
1.25$.
(For further details, see \citep{SN93J-2} and \citep{SN93J-4}.)}
\label{roriplot}
\end{SCfigure}

Our \uv~plane model allows a simultaneous fit of \thout\ and \thin\
when the resolution is sufficient.  For times later than 2.7~yr we
therefore obtained separate estimates for \thout\ and \thin, and can
study the evolution of the supernova's shell thickness.  We plot these
values of \thout\ and \thin\ in Figure~\ref{roriplot}, again scaled to
the nominal power-law expansion with $\theta \propto t^{0.8}$.  For
times between $t = 2.7$~yr and $t = 8$~yr the ratio $\thout / \thin$
does not change dramatically, being 1.29 on average (as we reported in
\citep{SN93J-2}).  However after $t \simeq 8$~yr, it can be seen that
the shell thickness increases systematically, with $\thout / \thin$
reaching $\sim$1.7 by the time of our latest observations at $t \simeq
15$~yr.  This increasing thickness of the shell is seen at different
frequencies and observing epochs, and is notably larger than the
observational uncertainties.  We argue in \citep{SN93J-4} that the
outer radius of our shell model likely corresponds well to the average
radius of the outer shock.  Although the relation of the inner radius
of our shell model to the reverse shock is less unambiguous, and is
likely somewhat model-dependent, we think that our results nonetheless
reflect a real increase in the ratio between outer and inner shock
front radii since $t \simeq 8$~yr.  A thickening shell is also seen in
simulations \citep{MioduszewskiDB2001}, although the thickening is not
as rapid as we have observed.  Our observations therefore imply
further deviations from self-similar evolution of the supernova, with
the outer radius continuing to increase in an approximately power-law
fashion, but with the average inner radius increasing more slowly,
perhaps as a result of instabilities growing inwards into the
unshocked ejecta.



\newcommand{\araa}{Ann. Rev. Astron. Astrophys.}
\newcommand{\aap}{Astron. Astrophys.}
\newcommand{\aapr}{Astron. Astrophys. Rev.}
\newcommand{\aaps}{Astron. Astrophys. Suppl. Ser.}
\newcommand{\aj}{AJ}
\newcommand{\apj}{ApJ}
\newcommand{\apjl}{ApJL}
\newcommand{\apjs}{ApJS}
\newcommand{\apss}{ApSS}
\newcommand{\baas}{BAAS}
\newcommand{\memras}{Mem. R. Astron. Soc.}
\newcommand{\memsai}{Mem. Soc. Astron. Ital.}
\newcommand{\mnras}{MNRAS}
\newcommand{\iaucirc}{IAU Circ.}
\newcommand{\jrasc}{J. R. Astron. Soc. Can.}
\newcommand{\nat}{Nat}
\newcommand{\pasj}{PASJ}
\newcommand{\pasp}{PASP}

\small
\bibliographystyle{JHEP}

\bibliography{mybib1}

\end{document}